**Attilio Sacripanti °*^ Antonio Pasculli “**
°ENEA (National Agency for Environment Technological Innovation and Energy)
*University of Rome II “Tor Vergata” Italy
^ FIJLKAM Italian Judo Wrestling and Karate Federation
“D’Annunzio University (Chieti) Italy

# Match Analysis an undervalued coaching tool

*“An Italian judo Federation contribution”*

Abstract
From a Biomechanical point of view, Judo competition is an intriguing complex nonlinear system, with many chaotic and fractals aspects,
It is also the test bed in which all coaching capabilities and athlete’s performances are evaluated and put to the test.
Competition is the moment of truth of all conditioning time, preparation and technical work, before developed, and it is also the climax of the teaching point of view.
Furthermore, it is the most important source of technical assessment.
Studying it is essential to the coaches because they can obtain useful information for their coaching.
Match Analysis could be seen as the master key in all situation sports (dual or team) like Judo, to help in useful way the difficult task of coach or best for National or Olympic coaching equips.
In this paper it is presented a short summary of the most important methodological achievement in judo match analysis.
It is also presented, at light of the last technological improvement, the first systematization toward new fields.
On the basis of technological development, Match Analysis could be a valuable source of four levels of information:

1st. Athlete’s Physiological data
2nd. Athlete’s Technical data
3rd. Athlete’s Strategically data
4th. Adversary’s Scouting

These new interesting ways opened by this powerful coaching help, are very useful for national team technical management.
In the last part of the paper, the analysis is focalized toward a, till now, misused information: “***Dromograms***” (athletes’ shifting paths) study as useful source of fighting habit of athletes.

**Attilio Sacripanti °*^   Antonio Pasculli “**
°ENEA (National Agency for Environment Technological Innovation and Energy)
*University of Rome II “Tor Vergata” Italy
^ FIJLKAM Italian Judo Wrestling and Karate Federation
“D’Annunzio University (Chieti) Italy

# Match Analysis an undervalued coaching help

*“An Italian judo Federation contribution”*

## Introduction

Italian Judo Federation FIJLKAM, member of the European Judo Union, now days made a big effort to develop new advanced methods for coaching help, coming into the telematics era.
Match Analysis systems are the new frontier of the technology, but the analysis of data competitions have a long and interesting life.
The birth of Match Analysis may refer to the single Athlete performance study for simple cyclic Sports.
It can be considered as extension from the “simple” before described case to two interacting athletes’ complex field, or again to more complex system of two teams mutually interacting.
Match Analysis can be considered as the master key in all situation sports (dual or team) like Judo, to help in useful way the difficult task of coaching, for National or Olympic coaching equips.
They started as notational analysis; made by Soviet Union and also by US and many other Countries at the end of seventy years but few people know that in that time already Japan Country showed a sort of futuristic vision made a very convincing and scientifically complete approach to Judo Match Analysis.
In the historic paper of Matsumoto, Takeuchi and Nakamura ***“Analytical studies on the contest performed at the all Japan Judo Championship Tournament”*** edited by Kodokan Scientific Bulletin in 1978 we can find a lot of information, very advanced for that time, with a clear both scientific and coaching vision of the whole competition problem. The summary content of this paper is shown:

***Execution of techniques***

*1st. number of times of executing techniques*
*2nd. decisive techniques (Kimari Waza)*
*3rd. number of executions of techniques and the issue of the contest*
*4th. type of techniques and the issue of the contest*

***Defence***

*1st. method of defence*
*2nd. method of defence in relation to body weight*
*3rd. method of defence in relation to different techniques*

***Left handed judoists***

*1st. the technique used and the method of clinching in relation to right or left*

***Method of use of the arena***

*1st. the distance of the movement of the judoists*
*2nd. various movement tendencies of the judoists*
*3rd. the area of the arena used*
*4th. the position of the judoists in relation to the inner or outer part of the arena when executing the techniques*
*5th. the position of the judoists on the arena when executing the decisive techniques*
*6th. outer side of the arena*
*7th. the use of a particular area in the arena by the judoists*

***Lost time***

**But this very important and complete paper was forgotten and became practically unknown.**
The following figures, taken by Matsumoto and co workers, show the analysis' results of the use of the Tatami area, performed by the Japanese experts.

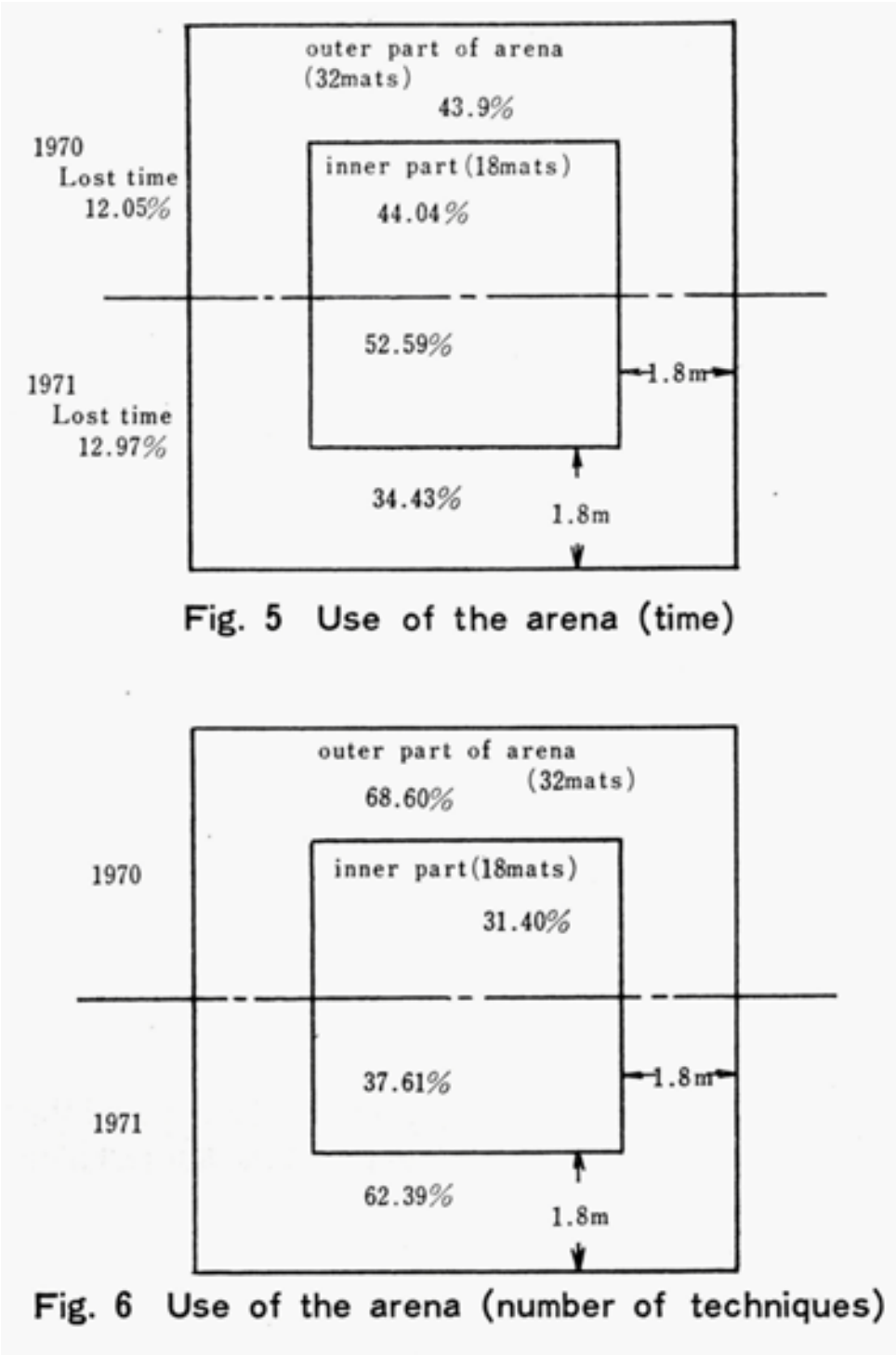

Fig. 5 Use of the arena (time)

Fig. 6 Use of the arena (number of techniques)

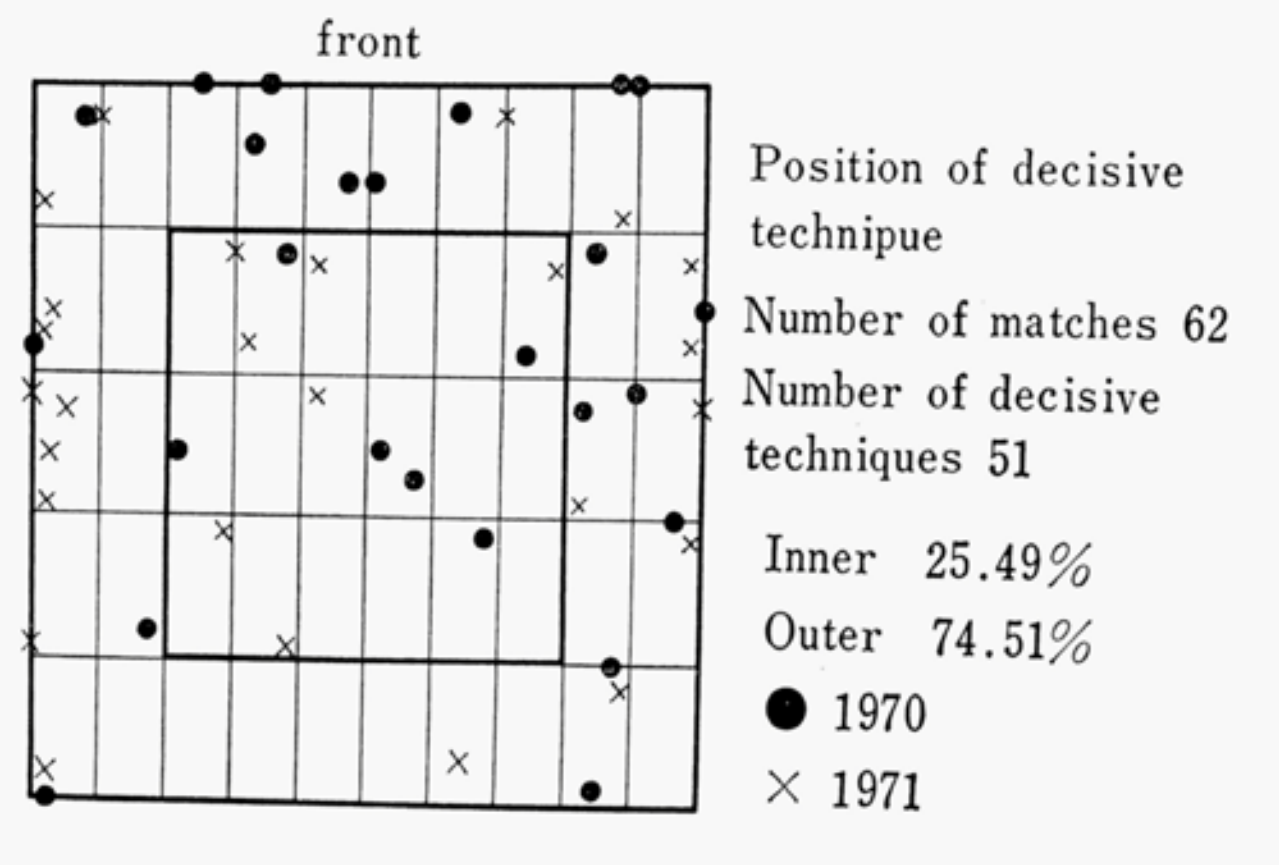

## I) The Computer Era

With the start of the computer era; the first Match Analysis advanced utilization in the world, like US and URSS was simply the same data sheets (for the notation before remembered) treated by more powerful statistical methods.
Match Analysis was executed by some Judo's national federation easily and speedily.
From year to year, Match Analysis was used with the advance in technology, for example: very small and portable high speed digital camera recording, at first with the only slow motion, now equipped with specific analysis software.
Other improvements are connected to the terrific increase in power and flexibility of hardware and software utilized.
In these times, Match Analysis was also developed for team sports mainly football, soccer and basketball, and these systems are today at the up data of the sport's world.
There are today two main utilization of video data collection:

a. *In Real Time*
b. *Off line*

The most advanced software for real time utilization come from volleyball, all other sports team or dual use match analysis systems mainly in off line studies.
We restrict ourselves at the off line application of judo. The reason is that judo match is short in time and this situation makes more difficult the utilization of real time analysis.
The data video obtained will be quantitatively useful only if the camera is calibrated. In other way, only qualitative analysis can be performed.
The most important problem, in the off line analysis case, is focalized in the saving time automatic treatment of data base content.

On the basis of the technological development, quantitative or pseudo-quantitative Match Analysis could be today a valuable source of four levels of coaching information:

***1st.*** ***Athlete's Physiological data***
***2nd.*** ***Athlete's Technical data***
***3rd.*** ***Athlete's Strategic data***
***4th.*** ***Adversary's Scouting***

These new interesting ways opened by this powerful coaching assist, are very useful for all judo operators.

***1st.*** In the Area of Physiological data, the quantitative Match analysis, well performed, could help in two main ways:

a. Gross evaluation of the more precise energy cost as input for conditioning
b. Accident prevention

***2nd.*** In the area of the Technical data, essentially qualitative Match Analysis could be a very useful source of information for the athletes' biomechanical enhancement, in six main ways:

a. Biomechanical evaluation of grips methods and fight.
b. Biomechanical improvement of throws techniques.
c. Biomechanical improvement of Standing-laying connection
d. Biomechanical improvement of Ne waza techniques.
e. Biomechanical improvement of defensive systems.
f. Weak point's analysis.

***3rd.*** In the area of Strategic data, qualitative-quantitative Match Analysis could be a very important coaching assist, in the study and the refinement of strategic plans adopted; by two ways:

a. Local Strategies.
b. Global Strategies.

***4th.*** The same qualitative data base of athletes could be studied from a different point of view, to gather information about most dangerous adversaries for each fighter, by the main utilization today:

a Adversary's scouting.

**II) The Statistical Approach**

Normally another result by video analysis is the statistical treatment of data.
This approach is focalized on obtaining some so called performance index, that are specific numbers or time functions that would give compact and clear information about fight conduction or specific athletes' population trends ( like most applied techniques, time of application, and so on ).
Most of this information, during the last fifteen years, has been produced by the Polish school (Cracow and Warsaw). Worthy of mention are the works of Sikorski, Sterkowicz, Maslej, or Boguszewski and Boguszewska, other come from Brazil Franchini and co workers, others from Spain, for example Castarlenas and Planas, other from Bosnia Herzegovina Kajimovic and so on.
The most useful application of this kind of approach is to try to forecast some tendential evolution of data. "*Forecasting*" is the process of estimation in unknown situations. It can refer to estimation of time series data to obtain a more general tendential prediction for the future.
However, Statistical approach is a very useful tool of technical qualitative tendential information; it must be seen as complementary, in the real Match Analysis approach.

**III) Considerations on video data**

Qualitative or quantitative video recordings and player tracking patterns in games are probably the most complex representations of sports movements that can be come across.
This is true not only for the movements of the body segments of one athlete, which sports biomechanists generally, focus on, but also for the movement patterns of the athletes as couple or team. Mathematical models based on biophysical laws can give a sound theoretical basis to the analysis, which can otherwise become data-driven; most of these models are too far removed from coaching to be of practical use.
Good quantitative analysts need a sound grasp of techniques or movement interactions involved in a specific activity. However, these sequences are complex because they contain much information.
It is often beneficial for the qualitative or quantitative movement analyst to look at simpler representations of movement patterns.
In Judo fight analysis, sequences of still video frames must be used in analysing players' "complete movements" and "interactions". Interaction in judo throws is connected to the shifting speed of the Couple of Athletes and founded on two main physical principles application:

1. ***Couple of forces application***
2. ***Physical lever application***

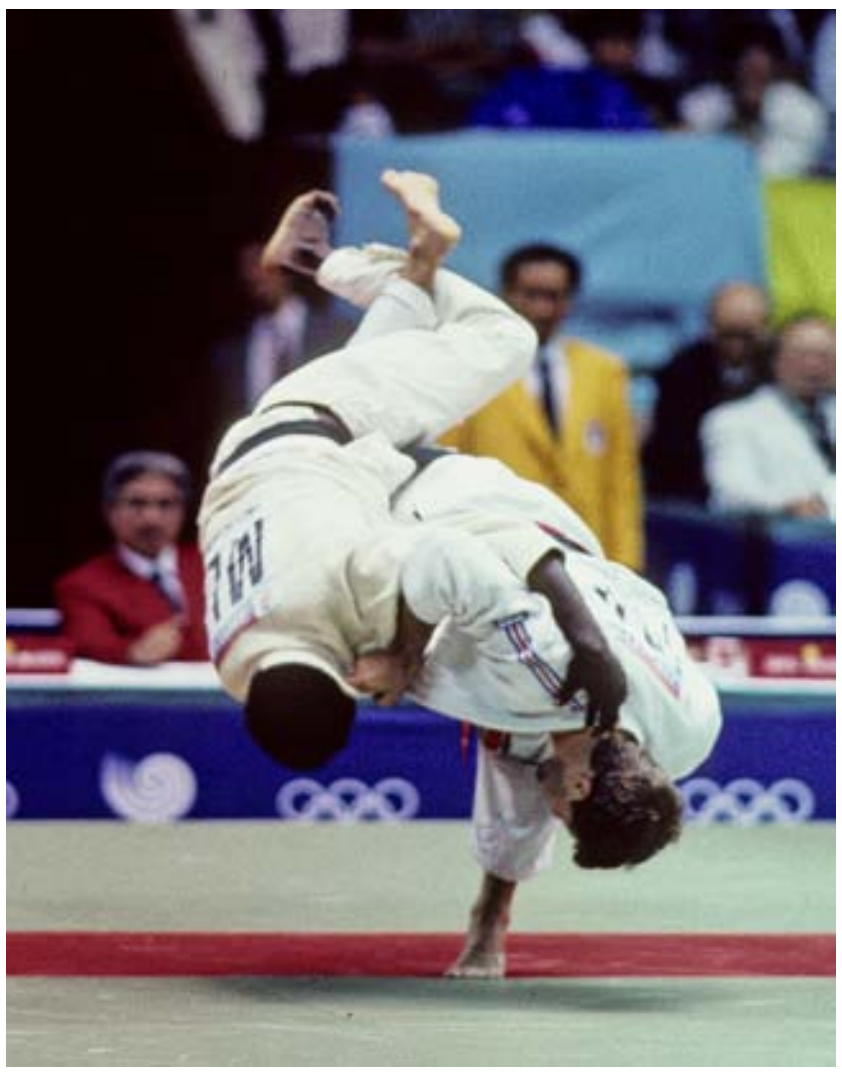

*Couple of Forces application*

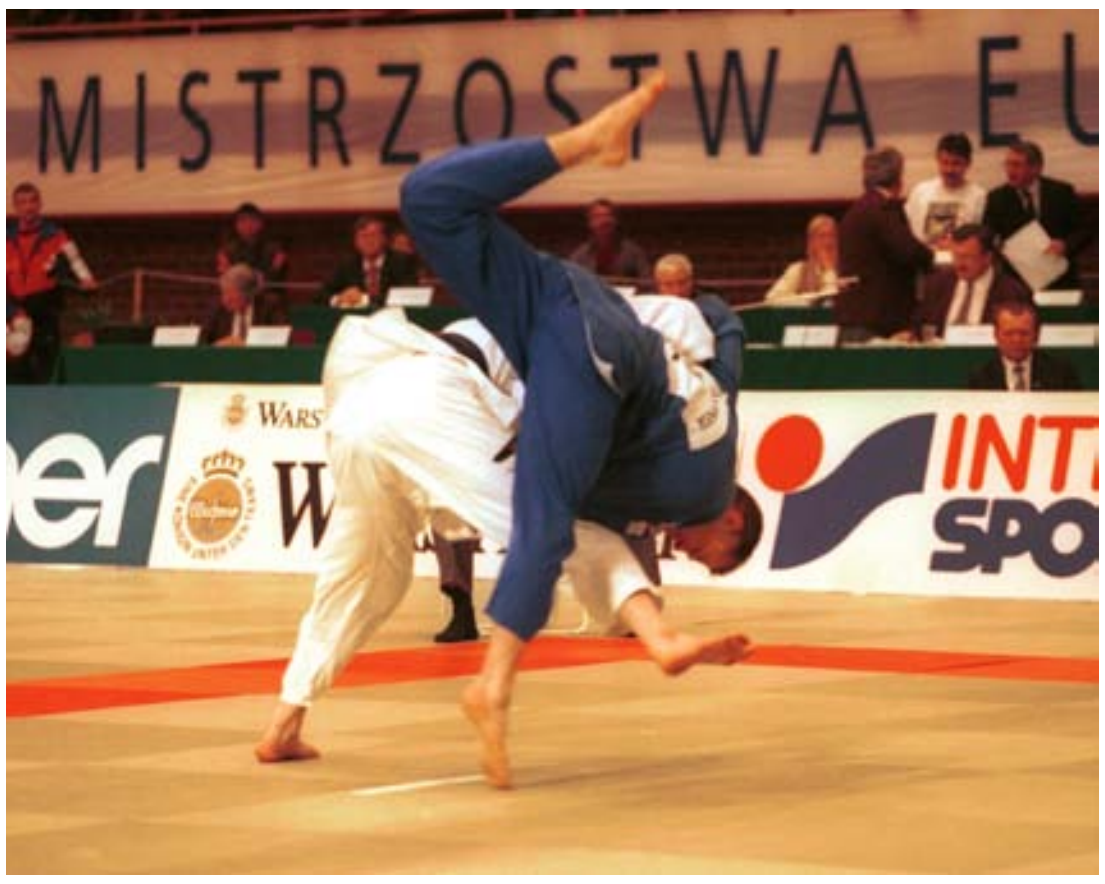

*Physical Lever Application*

Complete movements are connected to the today's most analyzed fighting phase *The Kuzushi – Tzsukuri movements* as it is possible to find in the works of Imamura and Komata and co workers.

**IV) Handling Video Data Base**

The video stream automatic handling is one of the most advanced boundaries of researches in many fields: engineering, computer science, physics, Image analysis and treatment, and so on.
Today, Match Analysis data base is performed automatically by means of special data mining algorithm or, more sophisticated systems that singles out tactical or specific strategic actions from the video recorded.
The most sophisticated software is based on the HHMM (Hidden Hierarchical Markov's Models) or on the Neural Networks with application of Bayesian Statistics.
They are able to single out specific determined sequences connected to strategy.
The next figure displays a Bayesian structure to find a fight interaction between two athletes.

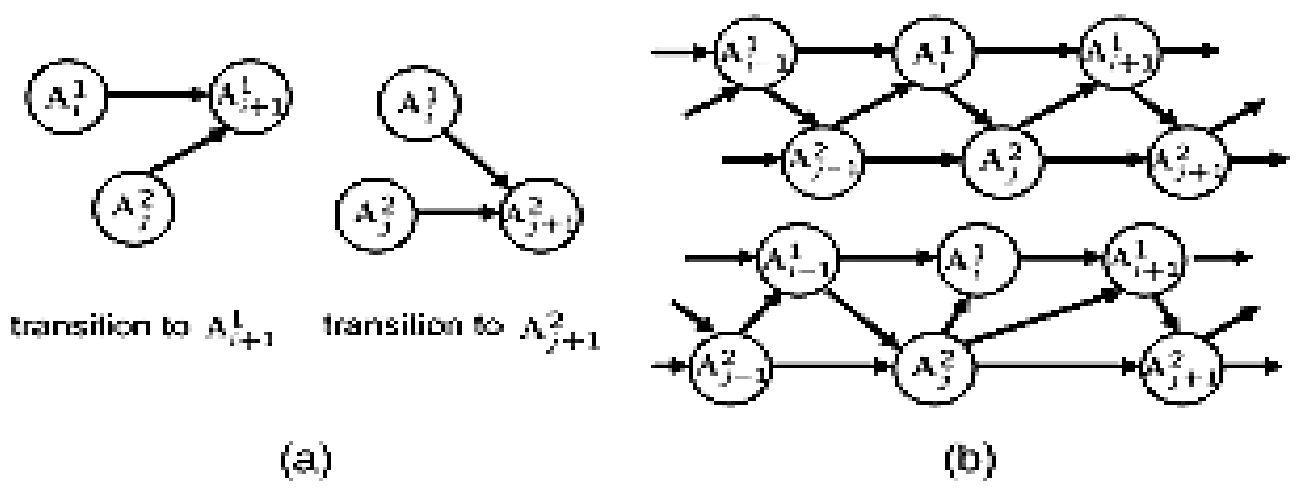

Fig. 8. Bayesian network is adopted to capture the interactions between two players. The building blocks shown on the left are combined dynamically to yield a variety of network structures. Two examples are shown on the right. (a) Building blocks. (b) Examples.

This is a huge problem for team sports, and until now these previous mentioned methods have never been applied to dual fighting sports like judo.
Video Data base are considered by coaches and Federations very treasure in the teaching area for elite athletes, in the light of the first paragraph classification.
Under this specific technical vision, the structure of these data base must be carefully structured and analyzed to extract from the video streams all important information.
Discussion about data base management is out of the goal of this paper, but today the correct management of the data base is the most important aspect of the question, because it is not only necessary for coaches and Federations to obtain full information , but it is also important to make sure that these information are error free. In the following figures are showed a judo data shots automatically optimized for similitude

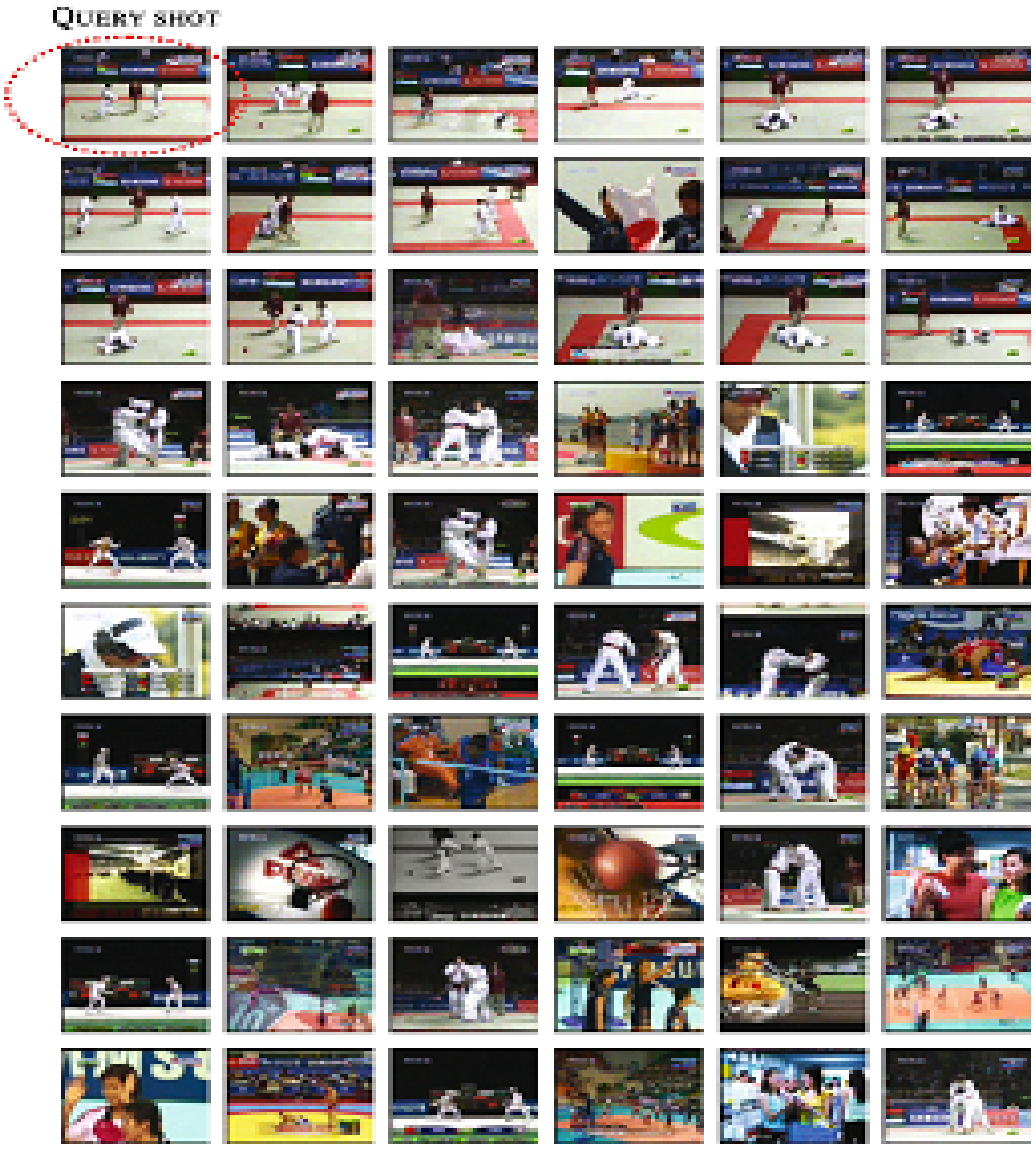

Fig. 3 Query results for a judo shot

## V) Athletes' Tracking Paths an underestimated practical tool

Player tracking patterns in games are a very useful tool for coaching both in team competitions, such as basketball and soccer, and in dual competitions, such as tennis and judo.
Normally tracking is based on the study of one point motion for example athletes' COM, or as in judo in the study of Couple of Athletes COM projection on the tatami.
The first analysis of this aspect of competition had been realized by Matsumoto and co-workers in the historical paper previously mentioned

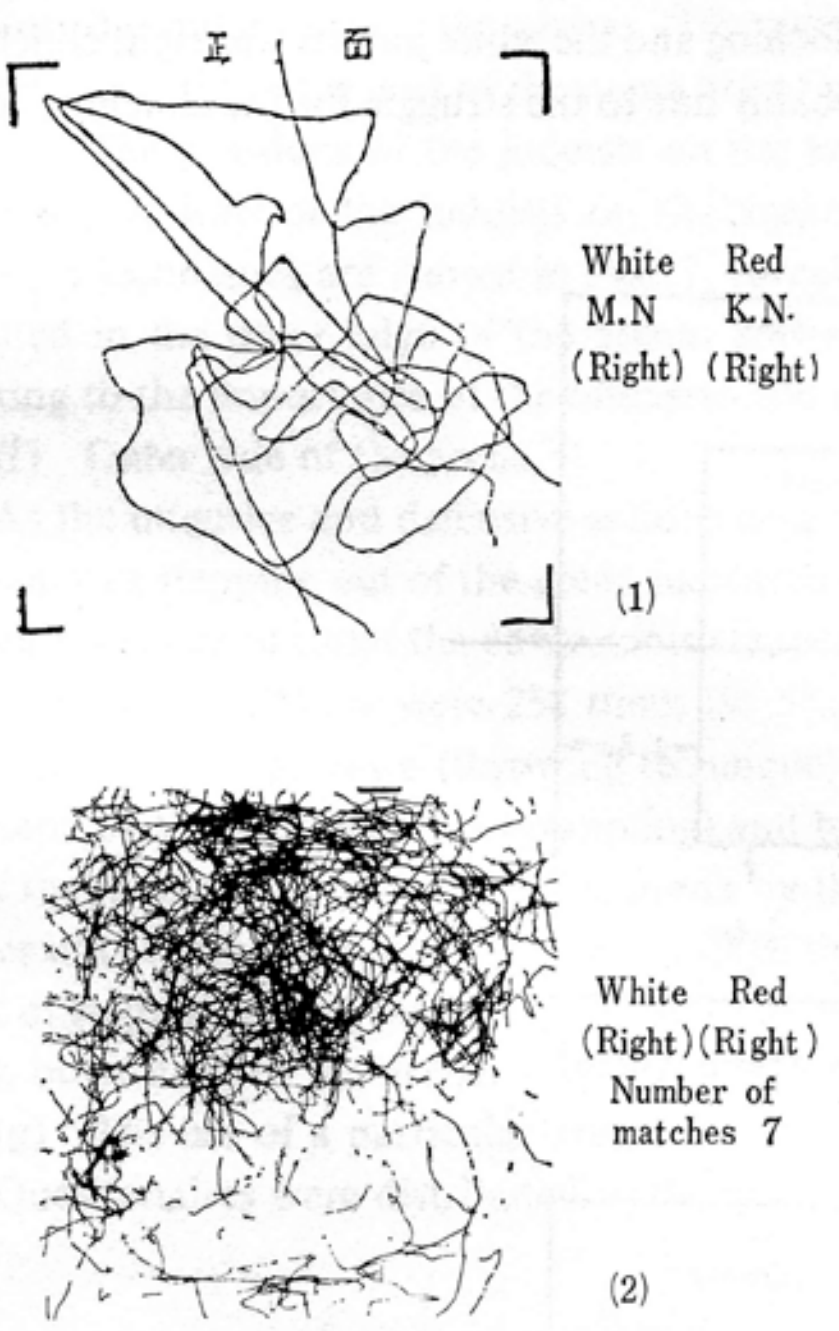

Fig. 3

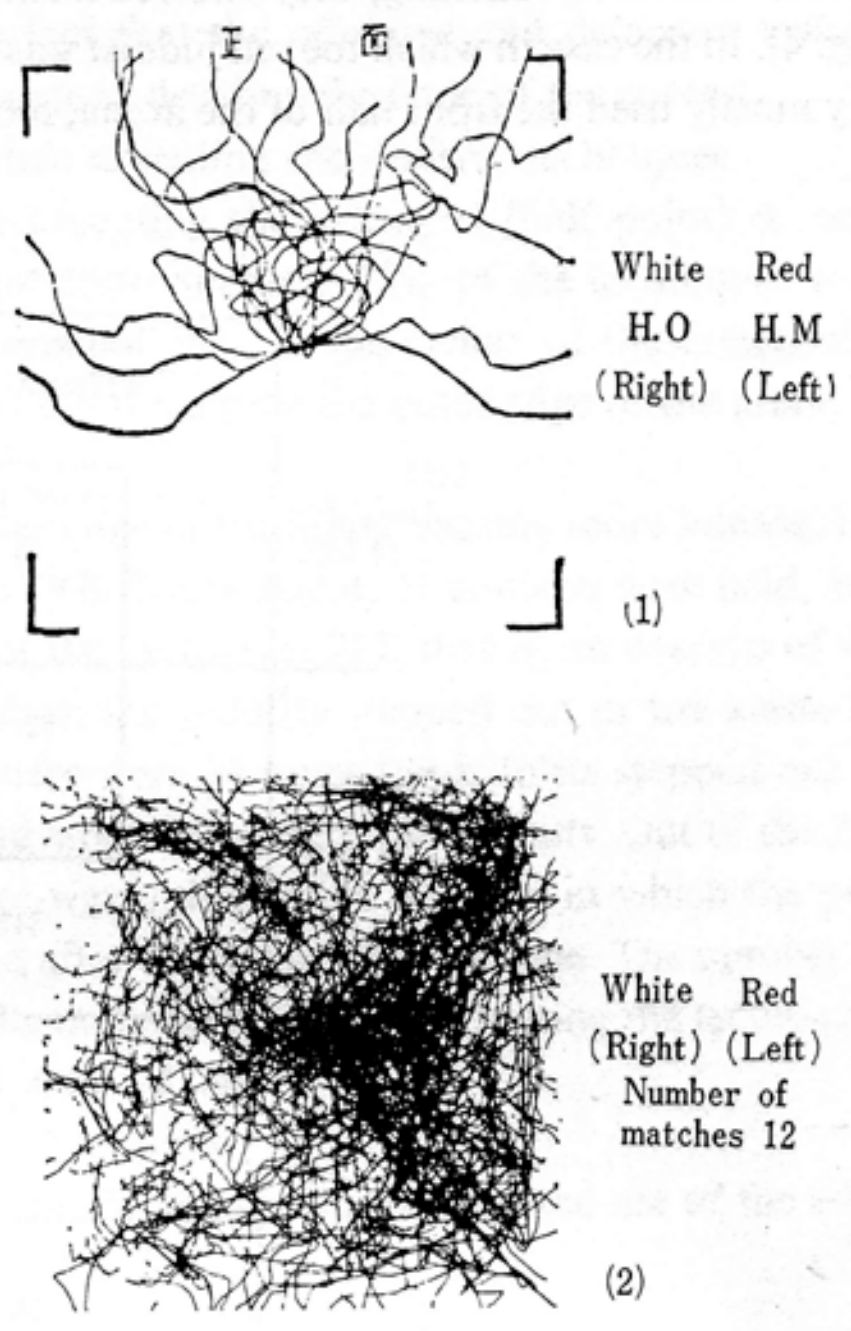

Fig. 4

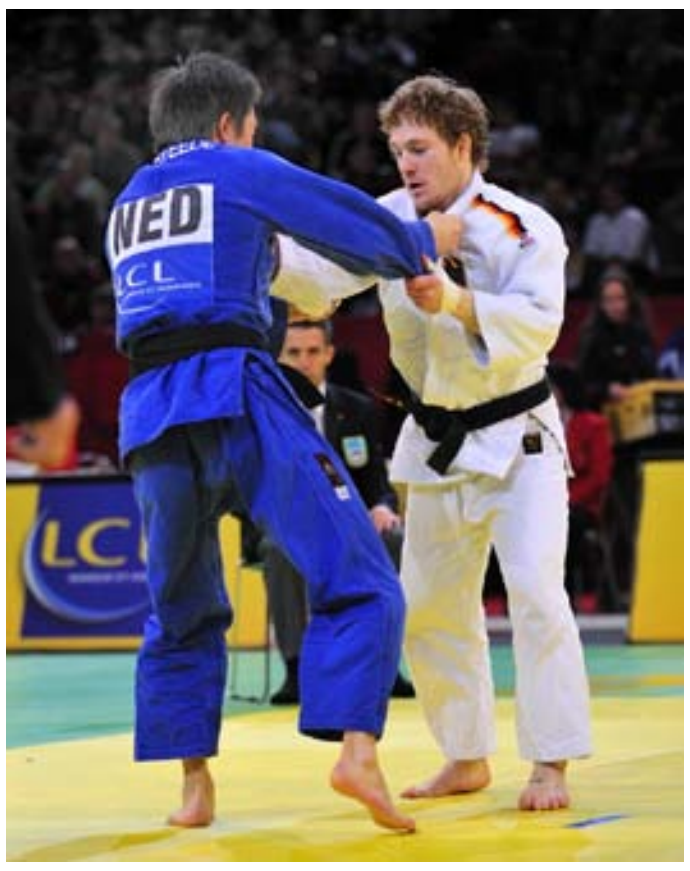

***Normal Classical Grips***

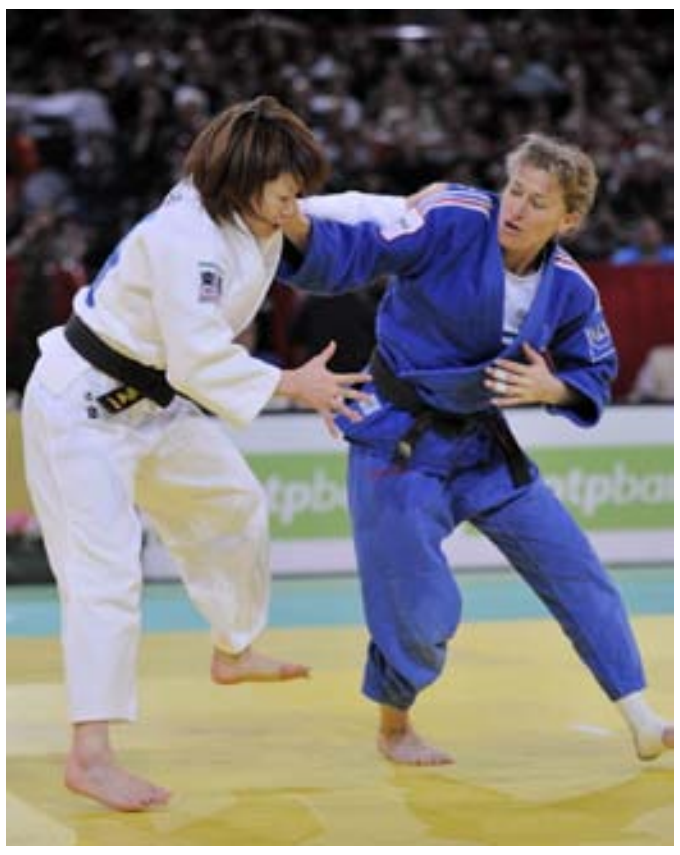

***Opposite Classical Grips***

But the analysis of the paths was only connected to the grips position and Matsumoto and co-workers stated that: "*the judoists' direction of movements, especially in relation to the form of clinching was studied, and it was found that in contest between judoists with the same form of clinching, they revealed the tendency to circle the arena widely, Fig3 . In a contest between judoists with different form of clinching, they showed the tendency to move directly toward the outside of the arena Fig 4*" They obtained also the mean distance covered by judokas ( ***121,1 m***) and the mean shift velocity ( ***0.30 m/s***).
These were all the results obtained by the Japanese researchers, about shifting paths.

By then, authors can't find any other study on this specific aspect of competitions.
Are these analyses useless or less important in content for Match Analysis?
The authors believe: that this statement is not true, and that the content of trajectories analysis is full of interesting subtle information at light of most advanced treatment.

Today a video analysis system specifically for tracking must follow the next criteria:

- ***Low hardware cost*** - Analogical surveillance cameras, coupled with off-the-shelf DVD video recorders. DVD recorders can be placed at an easily accessible location.
- ***Flexibility*** - surveillance cameras can be fitted with many brands and types of inexpensive lenses.
- ***Off-line processing*** - videos are recorded during the match, and then transferred to personal computer after the recording. Off line processing provides a way for obtaining accurate and reliable data, as the processing results can be reviewed and corrected.
- ***Automated processing*** - advanced computer vision methods are employed for tracking the players, and the tracking is automatic.
- ***Operator supervision*** - the tracking process is initialized by human operator, and then supervised. Tracking errors are normally corrected by operator as well.

The motion patterns of the couple of athletes' system are a useful practical tool with hidden information inside.
The shifting patterns study could be source of very useful strategic data, but the price for extract the hidden information is a non trivial mathematical analysis of these special time series.
In fact, for what concern the tracking trajectories one of the authors demonstrated 20 years ago, that the shifting paths of the couple of athletes COM projection must be considered belonging to the class of Brownian motions.
In the fractional Brownian motion (fBm) approach, initially presented by Mandelbrot and van Ness in 1968, any time series can be considered a combination of deterministic and stochastic mechanisms.
The concept developed through fBm is, indeed, a generalisation of the Einstein's work, which showed that a stochastic process is characterised by a linear relationship between mean square displacements $< x^2>$ and increasing time intervals $t$, in formula:

$$\langle x^2 \rangle = 2D\Delta t \qquad (1)$$

The general principle of the fBm framework is that the aspect of a trajectory, expressed as a function of time, may be quantified by a nonfinite integer or fractional space dimension, hence providing a quantitative measurement of evenness in the trajectory.
It is possible to write in mathematical form:

$$D_t^{\alpha}\left[X(t)\right] - \frac{X(0)}{\Gamma(1-\alpha)} t^{-\alpha} = \xi(t) \qquad (2)$$

The first term is a fractional derivative, the second is connected to the initial condition of the process, and the third is always the random force acting on the COM.
In this case is important to know the mean square displacement of the point:

$$\left\langle \left[X(t) - X(0)\right]^2 \right\rangle = \frac{\langle \xi^2 \rangle}{(2\alpha-1)\Gamma(\alpha)^2} t^{2\alpha-1} \propto t^{2H} \qquad (3)$$

From this expression it is possible to understand that we are in presence of different diffusion processes, identified by the Hurst parameter.
In particular this parameter is time independent and it describes the fractional Brownian motion with anti-correlated samples for $0<H<1/2$ and with correlated samples for $½<H<1$.
If H is = to ½ we can speak of pure Brownian motion.

**V.1) Athletes shifting in computational Biomechanics**
***Numerical Evaluation***
The aim of this evaluation is to formulate a numerical strategy to evaluate reasonable athlete's paths, linked, in a simplified way as a first step, to the previous discussed theory.
The fundamental assumptions are introduced and described through the following points:

a. the two athletes are supposed to be located at the two ends of a bar;
b. the bar can rotate around its middle point, considered as the barycentre of the *athletes couple*;
c. the length of the bar oscillates in a *harmonic-random* way;
d. The centre of the bar (barycentre of the couple) moves by a *random motion* like.

Thus, the motion of the two athletes is simulated by a *random motion* of a "*pulsating*", "*spinning*" *circle*, with a mass equal to the sum of the athletes' masses.
The variable diameter depends on both their gripping arms length and their shoulder "thickness".
In all the following numerical simulations, a total 140 kg weight and a 0.2-0.8 m diameter range are assumed.
From the previous discussion, it follows that the global motion of a single athlete is assumed to be the vectorial composition of three different elementary movements:
- A *random motion* of the centre of the "*athlete's couple*",
- A *random rotation* around the centre;
- A *random oscillation* towards the centre.
It worth to observe that all the three elementary motions are characterized by a *random behaviour*.
The motion of the "*couple centre*" is characterized, at each step, by a direction *uniformly randomly chosen* (on 360 grades). Along the chosen direction a *displacement* is calculated supposing a *rectilinear uniformly accelerated* motion. The *time length* of the motion is evaluated, again, by a *random number generation approach*. In a more detailed way: the displacement of the *couple centre*, along the already chosen direction, is evaluated through the following equation ( Sacripanti):

$$m\frac{d\mathbf{v}}{dt} = -\beta\mathbf{v} - m\frac{(\mathbf{v}-\mathbf{v}_a)}{\tau} + \mathbf{P} - \mathbf{A}e^{-|x|/b} + \mathbf{L} \tag{4}$$

Where m is the total mass of the athletes couple, $\mathbf{v}$ is the vector velocity, $\tau$ is the relaxation time necessary to reach the "target velocity" $\mathbf{v}_a$, $\mathbf{P}$ a *push/pull force*, $\mathbf{A}\exp(-|x|/b)$ is a global term related to the border line distance strategy, while $\mathbf{L}$ is the *random Langevin force*. This kind of force is introduced through the *random chosen of the direction* and the *random chosen* of how long is the "*couple centre*" displacement before the *next change of direction*. Thus the deterministic behaviour of this kind of motion is simulated by the solution of the following scalar equation:

$$m\frac{dv}{dt} = -\left(\beta + \frac{m}{\tau}\right)v + \left(\frac{mv_a}{\tau} + P - Ae^{-|x|/b}\right) \tag{5}$$

Whose solution is easily?

$$\Delta s = \frac{c}{a}\Delta t + \left(\frac{v_0}{a} - \frac{c}{a^2}\right)\left(1 - e^{-a\Delta t}\right) \tag{6}$$

Where $\Delta s$ is the total displacement of the "*couple centre*" displacement during the time step $\Delta t$,
$a = \frac{\beta}{m} + \frac{1}{\tau}$,
$c = \frac{v_a}{\tau} + \frac{P}{m} - \frac{A}{m} e^{-|x|/b}$

We assume the *time step is a random variable* $\Delta t_{rand}$ belonging to a Gaussian statistics with a mean time step $\Delta t_m$ and a variance $\sigma_{\Delta t}$.
Thus, we have considered the following expression for the selected time *random variable*:

$$\Delta t_{rand} = \Delta t_m + n_{\Delta t} \cdot \sigma_{\Delta t} \cdot G_norm \qquad (7)$$

Where $n_{\Delta t}$ indicates the total range of variability ( $n_{\Delta t} = 2$ for all the following numerical simulation). Then a *normal* "G_norm" Gaussian distributed stochastic variable ($\mu = 0$ and $\sigma = 1$) can be provided by the *Box and Muller* (1958) algorithm:

$$G_norm = \sqrt{[-2 \cdot \ln(Y1_{rand})} \cdot \cos(2\pi \cdot Y2_{rand}) \qquad (8)$$

Where $Y1_{rand}$ and $Y2_{rand}$ are two *independent* uniformly distributed random variables. That means that the intrinsic routine, related to the selected Compilator (RAND in FORTRAN 97), to generate random variables should be called. This subroutine *must be called twice* in order to obtain the two independent (pseudo-random) variables.
A simple uniform statistics is assumed as a first step, for a *random rotation* around the centre.
Thus, it was not implemented a more realistic dynamics introducing an *angular moment equation*.
The *random oscillation* towards the centre is simulated as a harmonic motion.

Future improvements will regard some reasonable assumptions in order to build an "objective function". The main criteria will be to select a function correlated to the *strategy* of the player around which, in a necessarily *randomly* way, a *tactic* function should be added. The *strategy* depends on the player characteristics.
Some numerical realizations of possible athletes paths, belonging to a statistical ensemble built up by the previous described approach have been carried up.
Fig. 1 shows how the athletes' movements are simulated by a *moving*, *spinning* and *pulsating* "*athletes couple circles*", shown in the red colour. The athletes are located at the ends of the reported red diameters, while the tracks of one of the two athletes are reported in black lines.

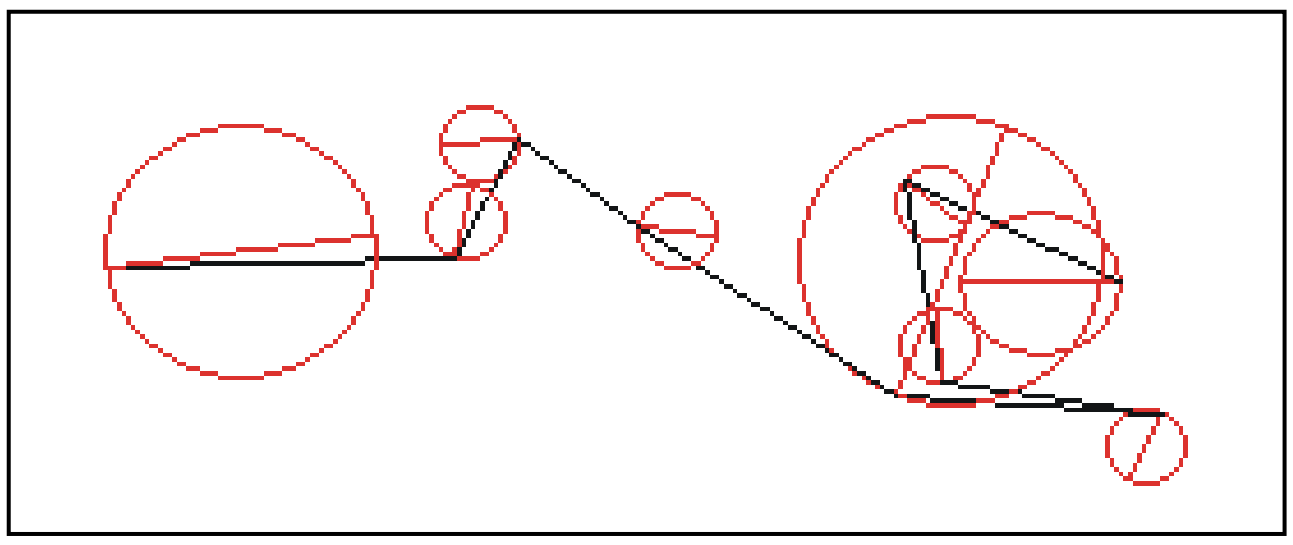

Fig. 1 Movements of the "*athlete's couple*"

In Fig.2 a single game, related to a possible realization belonging to a supposed statistical ensemble, is reported.

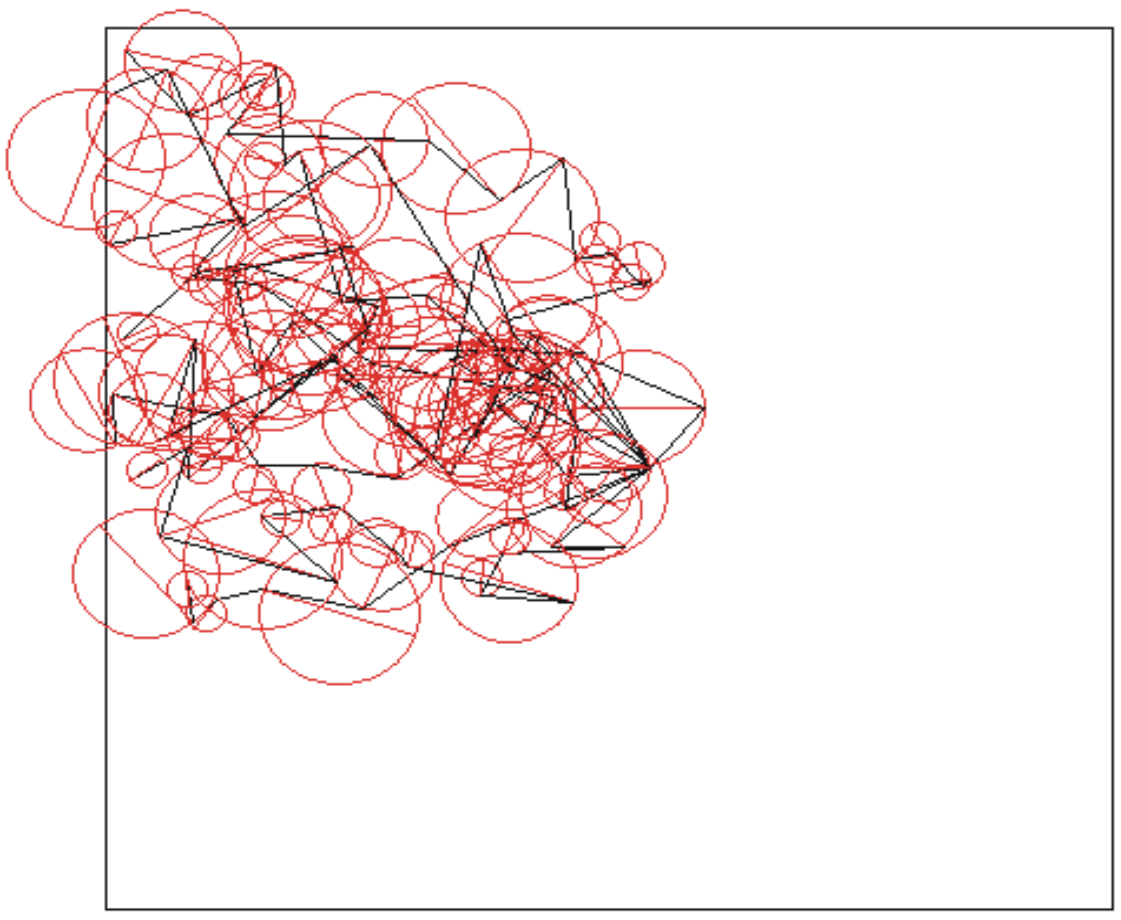

A single game "*realization*"

Fig.3 shows a superposition *realization* of the tracks of a single athlete, belonging to the *same statistical ensemble*, regarding *16 different virtual games*. The ensemble are built up supposing a completely symmetry.

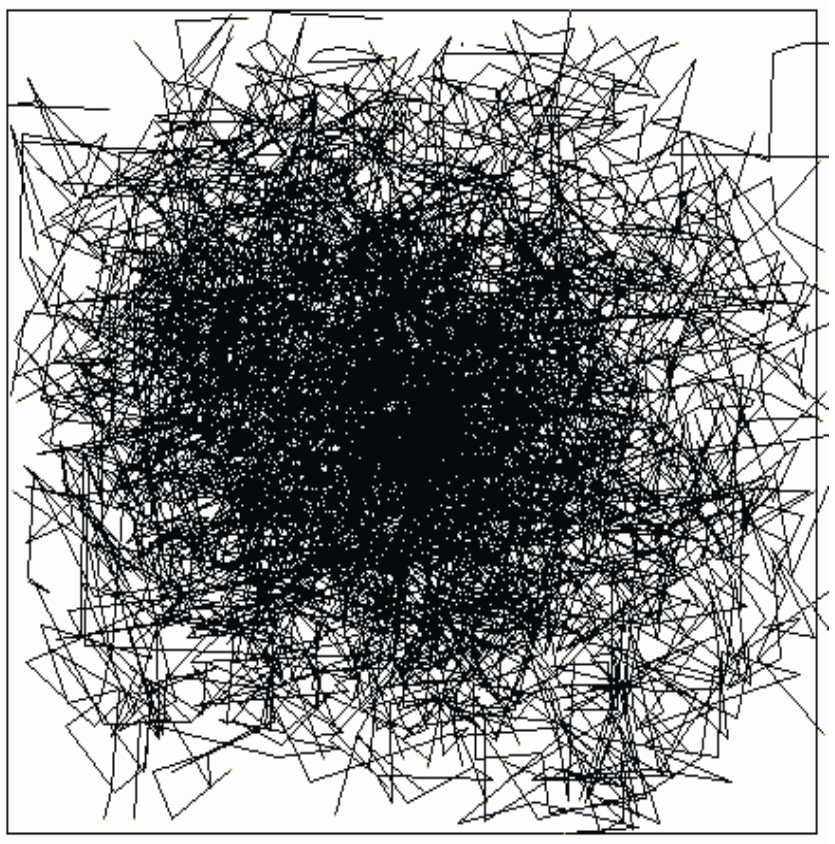

Fig.3 *Realizations* of 16 virtual games

In Fig.4a and 4b two different realizations of the tracks of a single athlete, belonging to the same statistical ensemble, regarding a single game, are reported.

The ensemble is built up supposing a smooth asymmetry along N-W direction.
Fig. 4c shows recorded tracks of an actual single game.
Also in Fig.5a and 5b two others different realizations of the tracks of a single athlete, belonging to the same statistical ensemble, regarding a single game, are reported.
This new ensemble is built up supposing a smooth asymmetry along S-N direction.
Fig. 5c shows recorded tracks of an actual single game.
In Figs. 6a, 6b, 6c; and Figs. 7a, 7b, 7c a superposition of both 7 and 12 single tracks, compared with a superposition of actual games are reported.

It is important to note that the changes of some parameters described before, allow describing different actual tracks. Another important point to note is the occurrence of *spatial permanence* of the tracks (Figs. 4a, 4b, 4c in particular). As the tracks depend on random movements, it seems that the *spatial permanence* could be related to the *Fractional Brownian* like motion (fBm) of the tracks (actual tracks as well).

Fig. 4a
**N-W** asymmetry
One game
A **1°** realization

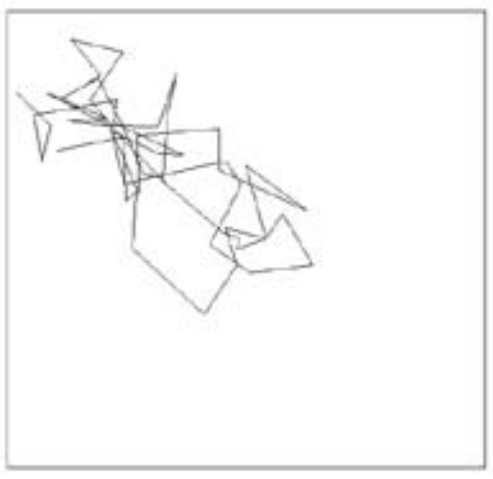

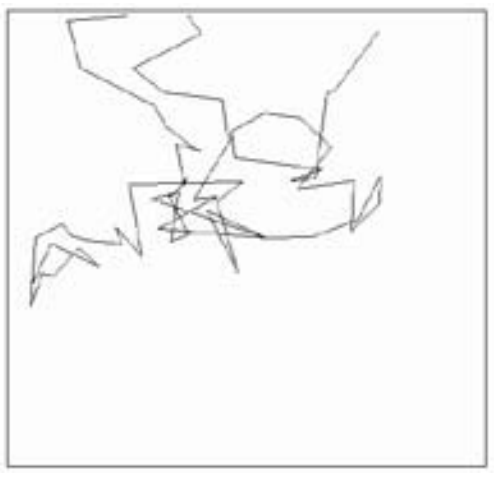

Fig. 5a
**S-N** asymmetry
One game
A **1°** realization

Fig. 4b
**N-W** asymmetry
One game
A **2°** realization

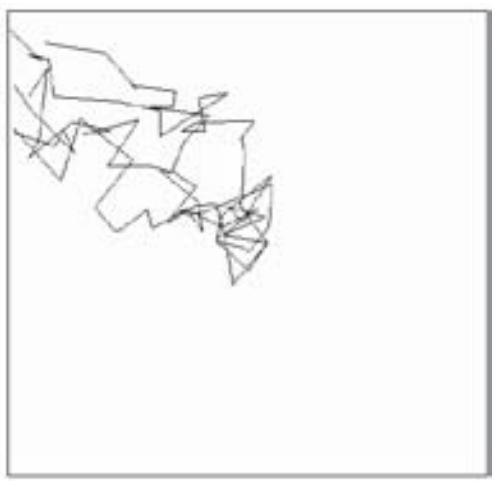

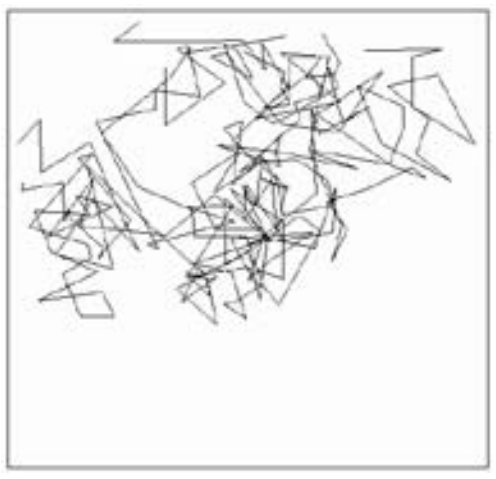

Fig. 5b
**S-N** asymmetry
One game
A **2°** realization

Fig. 4c
One game
**Experimental** tracks

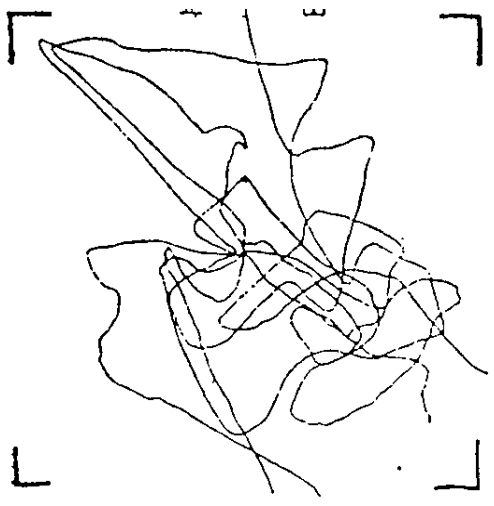

(1)

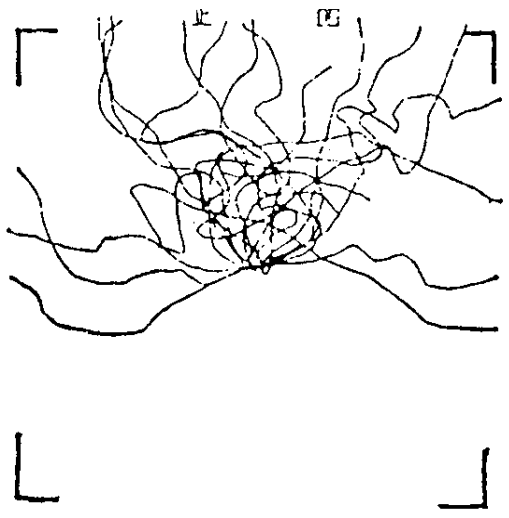

(1)

Fig. 5c
One game
**Experimental** tracks

Fig. 6a
7 games
**Experimental** tracks

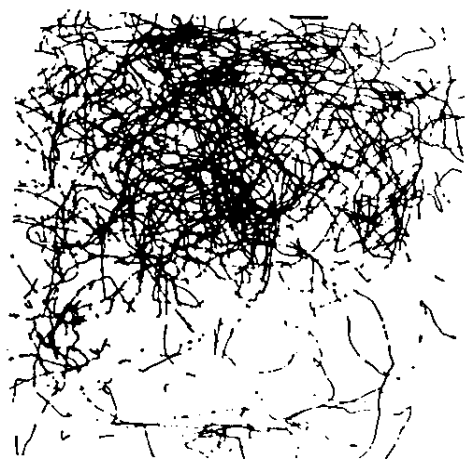

(2)

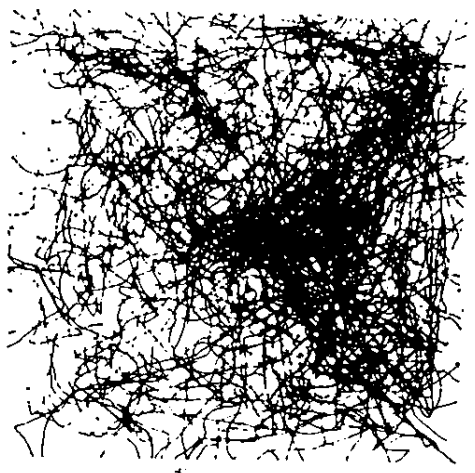

(2)

Fig. 7a
12 games
**Experimental** tracks

Fig. 6b
**N-W** asymmetry
7 games
A **1°** realization

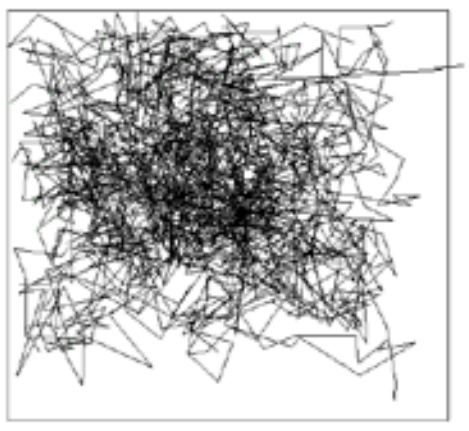

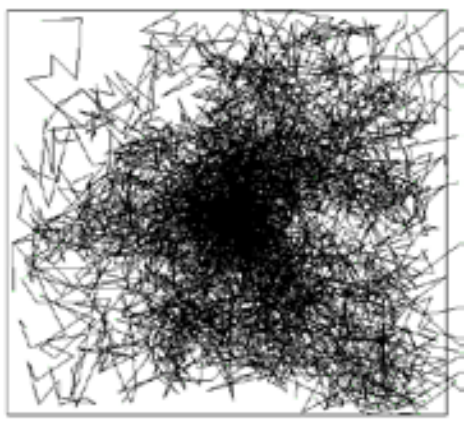

Fig. 7b
**N-E** asymmetry
12 games
A **1°** realization

Fig. 6c
**N-W** asymmetry
7 games
A **2°** realization

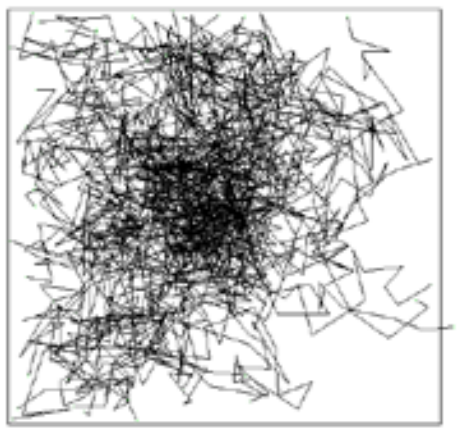

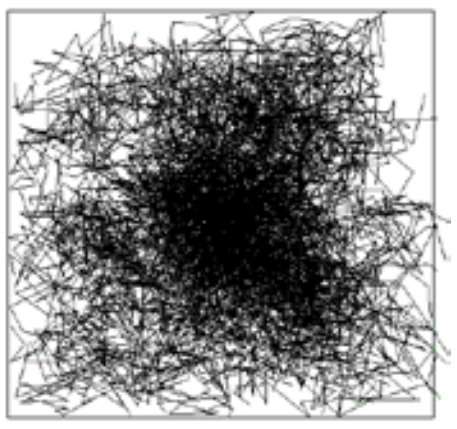

Fig. 7c
**N-E** asymmetry
12 games
A **2°** realization

**VI) Conclusion on strategic information obtainable for coaching**

Athlete's Tracks (***Dromograms***) are the evolution in time of the couple of Athletes COM projection on the tratami area.
Normally in the old Match Analysis each technical action and throw was considered belonging to a class of Markovian System, this means that it depends by the previous instant only, without correlation with the past movements.
A more advanced mathematical approach let able to overcome this conceptual limitation and mathematical simplification.
As we have seen before, an important feature of fBm modelling, for each fighter, is the presence of long-term correlations between past and future increments. This means that the system is not Markovian and then more similar to real situation.
This can be assessed by the scaling regimes.

***In this way <u>a fighting path can show</u>, if correctly analyzed, <u>when the fighter have a specific fighting strategy or not (random motion) during competition</u>.***

For example a median value of 0.5 for *H* indicates that there is no correlation, suggesting that the trajectory displayed a random distribution (Brownian motion).
On the other hand, if *H* differs from 0.5, a positive ($0.5 > H$) o r negative ($H < 0.5$) correlation with his fighting way can be inferred, indicating that a given part of initiative is under control.
Depending on how *H* is positioned, with respect to the median value 0.5, it can be inferred that the subject more or less controls the trajectory (and the fight evolution in time): the closer the regimes are to 0.5, the larger the contribution of stochastic processes (random attacks without strategy). In addition, depending on whether *H* is greater or less than the 0.5 thresholds, persistent (attacking) or antipersistent (defending) behaviours can be revealed, respectively. In other words, if the CM projection at a certain time is displaced towards a given direction, the larger probability is that it drifts away in this direction (persistent attacking behaviour) or in the contrary it retraces its steps in the opposite direction (antipersistent defensive behaviour).
Equality between these two probabilities indicates that there is not presence of a defined strategy in fighting, like simple random motion or stochastic process.
This information obtained by a pure "mathematical lecture" of trajectories; can be enhanced adding to the previous advanced mathematical lecture other Biomechanical fighting information like Grips form, Competition Invariants, Action Invariants, Attack useful polygonal surface, Direction of displacement, Time and position of gripping action, Throws "loci", Length or Amount of displacement, Medium Speed, and Surface Area Utilization and so on.
It is possible, with this added information, to obtain a lot of useful strategic information structured as tree and to treat this tree of information with opportune Data Mining algorithms to obtain a categorization of potentially effective strategic connection among shifting trajectories and other Biomechanical fighting information.
This information, ordered by importance or effectiveness, is useful for coaching and athletes as well.
This is one example of the more advanced information obtainable by this underestimated practical tool: Athletes' shifting patterns.

## VII) Bibliography

Franchini et al. ***Energy expenditure in different judo throwing techniques***. In: Yong Jiang, You-LianHong, Jin-Hai Sun. (Org.). Proceedings of 2008 Joint International Pre-Olympic Conference on Sports Science and Sports Engineering. 1 ed. Liverpool: World Academic Union, 2008, v. II, p. 55-60.),

Franchini et al ***A case study of physiological and performance changes in female judo players preparing for the Pan-American Games*** Rev. Bras.Cien. e Movida 9, N°2, 2001

Imamura R et.al ***"The theory of reaction resistance"*** Bulletin of the Association for Scientific Studies on Judo, Kodokan, Report XI, 2007

Kajmović et al. ***Differences analysis of situational efficiency performances between three level of judo competition for female seniors.*** Book of Abstracts.12th Annual Congress of the European College of Sport Science (pp.454-455). Jyvaskyla, Department of Biology of Physical Activity. Finland. 2007

Komata et al. ***A biomechanical Investigation of kuzushi of O soto gari in Kano Cup international competition***

Koshida et al. ***The common mechanism of anterior cruciate ligament injuries in judo : A retrospective.*** *Br J Sports Med 42* .2008;

Kruszewski et al. ***The evaluation of tactical and technical preparation of the senior medallists from 66 kg category participating in European Championships from 2004 to 2006.*** 2008

Janez Per ˇs, et al. ***Analysis of Player Motion in Sport Matches*** Computer science in sport –mission and method 2008

Matsumoto, Takeuchi and Nakamura ***"Analytical studies on the contest performed at the all Japan Judo Championship Tournament"*** Bulletin of the association for the scientific studies on judo, Kodokan Tokyo Report V-1978

Petrov ***Lutte libre et lutte Greco romaine*** FILA 1984

Sacripanti ***Biomeccanica del Judo*** ed Mediterranee 1988

Sacripanti ***Biomeccanica degli Sport di combattimento*** Ed FILPJK 1996.

Sacripanti ***Contest Dynamics: General Biomechanical Theory of contest sports*** see arXiv:0806.4069v1

Sacripanti et al ***Computational Biomechanics,Stochastic motion and team Sports*** see arXiv:0811.3659v1

Sacripanti ***Match Analysis I*** SDS 72- 2008 see also www.Scribd.com

Sacripanti ***Match Analysis II*** SDS 73- 2008 see also www.Scribd.com

Sikorski, W. et al. ***Structure of the contest and work capacity of the judoist***.: Proceedings of the International Congress on Judo "Contemporary Problems of Training and Judo Contest" 9-11 November, 1987, Spala-Poland, 1987. p. 58-65.

Snoek et al ***Multimodal Video Indexing: A Review of the State-of-the-art*** Multimedia tolls and application Springer 25, 2005

Solieman Osama K .***Data Mining in Sports: A Research Overview*** MIS master project 2006

Sterkowicz S., Maslej P.: ***„Analiza przebiegu walki judo na podstawie jej struktury czasowej-badania porównawcze".*** Sport Wyczynowy, 7-8; (1999)

Taesoo Kwon et al ***Two-Character Motion Analysis and Synthesis*** IEEE transaction and visualization in computer graphics 14, N°3, 2008

Tolga Esat Özkurt et al ***Principal component analysis of the fractional Brownian motion for 0 < h < 0.5*** IEEE-ICASSP 2006

Yuxin Peng et al ***OM-based video shot retrieval by one-to-one matching*** Multimed. Tool Appl. 34, 249-266, 2007

Xingquan Zhu et al ***Video Data Mining: Semantic Indexing and Event Detection from the Association Perspective*** IEEE transaction on knowledge and data engineering 17, N°5, 2005